\documentclass[conference]{IEEEtran}

\IEEEoverridecommandlockouts

\usepackage{cite}
\usepackage{amsmath,amssymb,amsfonts}
\usepackage{algorithmic}
\usepackage{parcolumns}
\usepackage{graphicx}
\usepackage{textcomp}
\usepackage{xcolor}
\usepackage{gensymb}
\usepackage{tabularx}
\newcommand\numberthis{\addtocounter{equation}{1}\tag{\theequation}}
\def\BibTeX{{\rm B\kern-.05em{\sc i\kern-.025em b}\kern-.08em
		T\kern-.1667em\lower.7ex\hbox{E}\kern-.125emX}}

\begin{document}
	
	\bstctlcite{IEEEexample:BSTcontrol}
	
	\newcommand{\heading}[1]{\colchunk[1]{\hspace*{-\parindent}\textit{#1}}}
	\newcommand{\desc}[1]{\colchunk[2]{#1}\colplacechunks}
	
	\title{Space-Constrained Arrays for Massive MIMO
		%\thanks{The work of M. Matthaiou was supported by EPSRC, UK, under grant EP/P000673/1.} %%% Funding source goes here
	}
	\author{\IEEEauthorblockN{%
			Chelsea L. Miller\IEEEauthorrefmark{1},
			Peter J. Smith\IEEEauthorrefmark{2},
			Pawel A. Dmochowski\IEEEauthorrefmark{1}
			%Harsh Tataria\IEEEauthorrefmark{3}, and
			%Andreas F. Molisch\IEEEauthorrefmark{4}
		}
		
		\IEEEauthorblockA{\IEEEauthorrefmark{1} School of Engineering and 
			Computer Science, Victoria
			University of Wellington, Wellington, New Zealand}
		\IEEEauthorblockA{\IEEEauthorrefmark{2}School of Mathematics and 
			Statistics, Victoria
			University of Wellington, Wellington, New Zealand}
		%\IEEEauthorblockA{\IEEEauthorrefmark{3}Department of Electrical and Information Technology,
		%	Lund University, Lund, Sweden}
	%	\IEEEauthorblockA{\IEEEauthorrefmark{4}Department of Electrical Engineering,
	%		University of Southern California, Los Angeles, CA, U.S.A.}
		\IEEEauthorblockA{e-mail:~\{chelsea.miller, peter.smith, pawel.dmochowski\}@ecs.vuw.ac.nz
			%,~harsh.tataria@eit.lth.se, and~molisch@usc.edu
		}
	}
	
	% \author{\IEEEauthorblockN{1\textsuperscript{st} Chelsea Miller}
	% \IEEEauthorblockA{\textit{School of Engineering and Computer Science} \\
	% \textit{Victoria University of Wellington}\\
	% Wellington, New Zealand \\
	% chelsea.miller@ecs.vuw.ac.nz}
	% \and
	% \IEEEauthorblockN{2\textsuperscript{nd} Pawel Dmochowski}
	% \IEEEauthorblockA{\textit{School of Engineering and Computer Science} \\
	% \textit{Victoria University of Wellington}\\
	% Wellington, New Zealand \\
	% \and
	% \IEEEauthorblockN{3\textsuperscript{rd} Peter Smith}
	% \IEEEauthorblockA{\textit{School of Mathematics and Statistics} \\
	% \textit{Victoria University of Wellington}\\
	% Wellington, New Zealand \\
	% \and
	% \IEEEauthorblockN{4\textsuperscript{th} Harsh Tataria}
	% \and
	% \and
	
	\maketitle
	
	\begin{abstract}
		We analyse the behaviour of a massive multi-user MIMO (MU-MIMO) system comprising a base station (BS) equipped with one of five different antenna topologies for which the spatial aperture is either \textit{unconstrained}, or \textit{space-constrained}. We derive the normalized mean interference (NMI) with a ray-based channel model, as a metric for topology comparison in each of the two cases. Based on the derivation for a horizontal uniform rectangular array (HURA) in \cite{JSTSP}, we provide closed-form NMI equations for the uniform linear array (ULA) and uniform circular array (UCirA). We then derive the same for a vertical URA (VURA) and uniform cylindrical array (UCylA). Results for the commonly-considered unconstrained case confirm the prior understanding that topologies with wider azimuth footprints aid performance. However, in the space-constrained case performance is dictated by the \textit{angular resolution} afforded by the topology, particularly in elevation. We confirm the behavioural patterns predicted by the NMI by observing the same patterns in the system SINR with minimum mean-squared error (MMSE) processing.
	\end{abstract}
	
	%%\begin{IEEEkeywords}
	%%%%%%  NOT IN A CONFERENCE PAPER
	%%\end{IEEEkeywords}
	
	\section{Introduction}\label{intro}
	As a result of standardization activities \cite{3GPP}, massive MU-MIMO will become a key technology in next-generation cellular systems. Antenna topology and spacing are key design elements in both microwave and millimetre-wave frequency bands. A large number of works examining the effects of antenna spacing focus on a specific topology, for example a ULA \cite{masouros_space-constrained_2015, tataria_uplink_2017}, or HURA \cite{Biswas2016}. Many others which compare topologies, for example \cite{aslam_performance_2019}, do not impose space constraints.
	%, usually chosen to be one half-wavelength, as smaller spacings are known to increase antenna correlation.
	
	Such comparisons can be misleading for two reasons. Firstly, in the unconstrained case topologies with larger spatial apertures in the azimuth domain are advantageous. Angular variation of incident rays is greatest in the azumth plane, hence larger azimuthal apertures are known to increase spatial diversity within the channel, improving performance \cite{aslam_performance_2019}. Secondly, larger apertures in the unconstrained case result in smaller antenna spacing in the constrained case, increasing antenna correlation and harming performance. Therefore, antenna configurations should be considered within constrained apertures in the interest of both practicality and fairness of comparison.
	
	With this aim, we focus on a metric which we refer to as the normalised mean interference (NMI) between two arbitrary users. This has been shown in \cite{JSTSP} to additionally serve as an indication of the ergodic cell-wide channel correlation, and performance with zero-forcing (ZF) and MMSE processing.
	%Some examples in literature, such as \cite{roy_channel_2018}, examine similar metrics by simulation.
	%metrics, such as  which simulates the NMI for a ULA, VURA, and UCylA with fixed antenna spacing.
	Closed-form expressions for similar metrics are derived and analysed under space constraints in \cite{masouros_space-constrained_2015,wu_favorable_2017,DanielAltamirano2016} for a ULA, HURA, and/or UCirA.
	% in \cite{wu_favorable_2017} for a ULA, HURA, and UCirA, and in \cite{DanielAltamirano2016} for a ULA and HURA.
	The authors of \cite{DanielAltamirano2016} additionally simulate the NMI for a VURA and a cubic array. The primary limitation of each of the closed-form results in \cite{masouros_space-constrained_2015,wu_favorable_2017,DanielAltamirano2016} is the assumption that the angles of arrival (AoA) of the incoming rays are uniformly distributed within a given angular spread. Measurements at 2.53 GHz presented in \cite{sangodoyin_cluster_2018} demonstrate that angles are more accurately modelled using a clustered ray-based model with Gaussian or Laplacian distributed cluster centroids and Laplacian subray offsets. The assumption of uniform AoAs severely underestimates the correlation and inter-user interference in the channel. This is illustrated in \cite{JSTSP} and \cite{chelsea_icc19}, which derive equations for the NMI for a HURA and a ULA with arbitrary angular distribution, examining both uniform and Gaussian/Laplacian angles. The analysis in \cite{JSTSP} also examines a VURA through simulation but does not provide any closed-form results for this topology.\footnote{Furthermore, the parameter settings used for the VURA placed it perpendicular to the examined ULA. This orientation puts the VURA at a disadvantage by setting it parallel to the angle around which the majority of the azimuth radiation is concentrated.} The generic results in \cite{JSTSP} and \cite{chelsea_icc19} have yet to be examined under space constraints.
	
	We build on and improve the work in \cite{JSTSP} and \cite{chelsea_icc19} by using the NMI to compare five topologies under space constraints with a generic ray-based channel model. More specifically:
	
	%CONTRIBUTIONS:
	\begin{itemize}
		\item We derive the NMI for a VURA, a UCirA, a UCylA, and a ULA;
		\item Using channel parameters derived from measurement in \cite{sangodoyin_cluster_2018}, we examine and compare the NMI of five topologies with and without space constraints;
		\item Based on the NMI trends, we confirm that in the unconstrained case, the larger azimuth apertures afforded by horizontal topologies are advantageous. However, in the space-constrained case all topologies achieve similar azimuthal resolution. Hence, arrays with a vertical dimension perform better due to the added angular resolution in elevation despite the fact that the elevation angle spread is usually much smaller than the azimuth spread.
	\end{itemize}
	%
	%
	%
	%%%%%%%%%%%% SYSTEM MODEL %%%%%%%%%%%%%%%
	\section{System Model}\label{systemmodel}
	A base station (BS) equipped with $M$ omnidirectional antennas lies at the centre of a circular cell of radius $r$ and receives uplink communication from $L$ single-antenna users (UEs) positioned uniformly randomly within the cell, outside of the exclusion radius $r_0$ around the BS. We consider a single-cell system with perfect CSI at the BS to simplify analysis, providing an upper bound on performance which is suitable for the purpose of topology comparison. We assume a %clustered\footnote{All analysis provided for the NMI can be adapted to a non-custered ray-based model by setting either $C=1$ or $S=1$, or both for the LoS case.} 
	ray-based model, motivated by measurements in \cite{sangodoyin_cluster_2018}, where the $M \times 1$ channel from a user $l$ to the BS, $\mathbf{h}_l$, is a summation of $S$ scattered subrays in each of $C$ scattering clusters:
	\begin{align}\label{eq:channel_model}
	\mathbf{h}_l = \sum_{c\in \mathcal{C}^{(l)}}\sum_{s = 1}^{S}\gamma_{c,s}^{(l)}\mathbf{a}\left(\phi_{c,s}^{(l)},\theta_{c,s}^{(l)}\right).
	\end{align}
	The ray coefficient $\gamma_{c,s}^{(l)} = \sqrt{\beta_{c,s}^{(l)}}\exp(j\Theta_{c,s}^{(l)})$ contains the power, $\beta_{c,s}^{(l)}$, and uniformly distributed phase, $\Theta_{c,s}^{(l)} \sim \mathcal{U}[0,2\pi]$, of subray $s$ within cluster $c$ of the channel for user $l$. We model the ray powers $\beta_{c,s}^{(l)}$ as a fraction of the total link gain for user $l$ such that $\sum_{c = 1}^{C}\sum_{s = 1}^{S}\beta_{c,s}^{(l)} = \beta^{(l)}$. We utilise the classical path-loss and shadowing equation such that $\beta^{(l)} = AX_l\left(d_l/d_0\right)^{-\Gamma}$,	for a user $d_l$ meters from the BS, where $A$ is a unitless attenuation constant representing the average attenuation at reference distance $d_0$ without shadow fading, $10\log_{10}(X_l) \sim \mathcal{CN}(0,\sigma_\textrm{sf}^2)$ models the effects of shadow fading, and $\Gamma$ is the path-loss exponent.

	The steering vectors, $\mathbf{a}(\phi_{c,s}^{(l)},\theta_{c,s}^{(l)})$, are functions of $\phi_{c,s}^{(l)}$ and $\theta_{c,s}^{(l)}$, the ray's azimuth angle of arrival (AAoA) and elevation angle of arrival (EAoA), respectively, and are defined in Sec.~\ref{s:analysis}. We measure the AAoA as the angle between the incoming ray and the $x$-axis in the azimuth $x$-$y$ plane. The EAoA is measured as the the angle between the ray and the $z$-axis. We define broadside as $\phi = 0$ in azimuth and $\theta = \pi/2$ in elevation. We implement a clustered ray-based model wherein each AAoA, $\phi_{c,s}^{(l)} = \phi_c^{(l)} + \Delta^{(l)}_{c,s}$, arises from a cluster central angle $\phi_c^{(l)}$ and a subray offset $\Delta_{c,s}^{(l)}$. Similarly, $\theta_{c,s}^{(l)} = \theta_c^{(l)} + \delta_{c,s}^{(l)}$.
	%\subsection{Cluster Sharing}\label{s:cluster_sharing}
\section{Normalised Mean Interference}
	The interference power between two distinct users with channels $\mathbf{h}_{l}$ and $\mathbf{h}_{l'}$ is given by $|\mathbf{h}_{l}^\textrm{H}\mathbf{h}_{l'}|^2$. Using the channel model in \eqref{eq:channel_model}, in \cite{chelsea_icc19} we simplify the NMI (denoted $\kappa$) to
	\begin{align*}
\kappa	&=\mathbb{E}_{\Theta,\phi,\theta}[|\mathbf{h}_{l}^\textrm{H}\mathbf{h}_{l'}|^2]/[M^2\beta^{(l)}\beta^{(l')}]\\
	&= \mathbb{E}_{\phi,\theta}[|\mathbf{a}^\textrm{H}(\phi_{c,s}^{(l)},\theta_{c,s}^{(l)})\mathbf{a}(\phi_{c',s'}^{(l')},\theta_{c',s'}^{(l')})|^2]/M^2.\numberthis\label{eq2}
	\end{align*}
	where $\mathbb{E}_{\Theta,\phi,\theta}[\cdot]$ is the mean over phases, AAoAs, and EAoAs. 	As  the ray angles and phases are independent and identically distributed, $\kappa$ in \eqref{eq2} can be written as
	\begin{align*}
	\kappa = \frac{1}{M^2}\sum_{m=0}^{M-1}\sum_{m=0}^{M-1}|\mathbb{E}_{\phi,\theta}[(\mathbf{a}(\phi,\theta))_{m}^{*}(\mathbf{a}(\phi,\theta))_{m'}]|^2.\numberthis\label{eq:kappa_raw}
	\end{align*}
	To verify the validity of the NMI as an indication of system performance with UL processing, we simulate the cell-wide SINR for user $l$ with MMSE processing given by \eqref{eq:MMSE_SINR} in \cite{tataria_revisiting}
	\begin{align*}
	\mathbb{E}_{\beta,\Theta,\phi,\theta}[\textrm{SINR}^\textrm{MMSE}_l] = \mathbb{E}_{\beta,\Theta,\phi,\theta}[\mathbf{h}_l^\textrm{H}(\mathbf{H}_l\mathbf{H}_l^\textrm{H} + \frac{1}{\rho}\mathbf{I}_M)^{-1}\mathbf{h}_l]\numberthis\label{eq:MMSE_SINR}
	\end{align*}
	where $\mathbf{H}_l = [\mathbf{h}_1, \dots \mathbf{h}_{l-1}, \mathbf{h}_l, \dots, \mathbf{h}_L]$ and $\rho$ is the UL SNR, the ratio of the symbol power to the noise power.
\section{Analysis of the NMI}\label{s:analysis}
	\newtheorem{lemma}{Lemma}
	\newtheorem{result}{Result}
	This section contains closed-form expressions for the NMI for all five topologies. We first provide a generic equation for the topologies which are confined to the $x$-$y$-plane (the HURA, ULA, and UCirA), then derive a second equation for those which utilise vertical antenna placement (the UCylA and VURA). Combining the appropriate equation and topology-specific parameters in Table \ref{table:topology_parameters} gives the NMI for each topology.
	%%%%%%%%%%%%%%%%%%%%%%%%%%%%%%%5%
	\subsection{HURA, ULA, UCirA}\label{ss:analysis_flat}
	Consider an HURA in the azimuth $x$-$y$-plane. $M$ antennas are arranged into $M_x$ rows separated by $d_x$ wavelengths along the $x$-axis, and $M_y$ columns separated $d_y$ wavelengths apart along the $y$-axis, where $M_xM_y=M$. The steering vector for a ray approaching at angle $\phi$ in azimuth and $\theta$ in elevation is
	\begin{equation}\label{eq:HURA_steering_vector}
	\mathbf{a}(\phi,\theta) = \mathbf{a}_x(\phi,\theta) \otimes \mathbf{a}_y(\phi,\theta).
	\end{equation}
	The $m_x^\textrm{th}$ element of the $M_x \times 1$ vector $\mathbf{a}_x(\cdot)$ is defined as \cite{wu_favorable_2017}
	\begin{align}\label{eq:ax_steering_vector}
	\left(\mathbf{a}_x(\phi,\theta)\right)_{m_x} = e^{j2\pi d_x (m_x-1) \sin{\theta} \cos{\phi}}
	\end{align}
	and the elements of the $M_y \times 1$ vector $\mathbf{a}_y(\cdot)$ are defined as
	\begin{align}\label{eq:ay_steering_vector}
	\left(\mathbf{a}_{y}(\phi,\theta)\right)_{m_y} = e^{j2\pi(m_y-1) d_y \sin{\theta}\sin{\phi}}.
	\end{align}
	From \cite{JSTSP}, we see that, for steering vectors defined as in \eqref{eq:HURA_steering_vector}, \eqref{eq:ax_steering_vector}, \eqref{eq:ay_steering_vector}, the NMI in \eqref{eq:kappa_raw} requires expectations of the form
	\begin{align*}
	\mathbb{E}_{\phi,\theta}[e^{j\sin\theta(z_1\sin\phi + z_2\cos\phi)}]
	 = \mathbb{E}_{\phi,\theta}[e^{j\sqrt{z_1^2 + z_2^2}\sin\theta\sin(\phi + A)}]\numberthis\label{eq:HURA_steering_vector_2}
	\end{align*}
	with $z_1 = 2\pi d_y(m_y - m_y')$ for $m_y,m_y'\in[1,M_y]$, $z_2 = 2\pi d_x(m_x - m_x')$ for $m_x,m_x'\in[1,M_x]$, $A = \tan^{-1}(z_2/z_1)$.
	
	The entries of the $M \times 1$ steering vector for a ULA situated along the $y$-axis are defined as in \eqref{eq:ay_steering_vector}; with this, \eqref{eq:kappa_raw} requires
	\begin{align*}
		\mathbb{E}_{\phi,\theta}[e^{j2\pi d (m-m') \sin\theta\sin\phi}]\numberthis\label{eq:ULA_steering_vector}
	\end{align*}
	for $m,m'\in [1,M]$. Many analyses of the ULA (\cite{chelsea_icc19, JSTSP, masouros_space-constrained_2015,wu_favorable_2017}) omit the $\sin{\theta}$ term in \eqref{eq:ULA_steering_vector}. This is equivalent to redefining the distribution of $\phi$ to account for variation in elevation. However, campaigns such as \cite{sangodoyin_cluster_2018} measure the true elevation and azimuth angles. We use \eqref{eq:ay_steering_vector} and hence \eqref{eq:ULA_steering_vector} to align with the use of measured data for $\phi$ and $\theta$. Note that \eqref{eq:ULA_steering_vector} is \eqref{eq:HURA_steering_vector_2} with $z_1 = 2\pi d(m-m')$, $z_2 = 0$, and $A = 0$.
	
	Finally, the steering vector entries for a UCirA with $M_r = M$ antennas spaced $d_r$ wavelengths apart in a circle in the $x$-$y$-plane are defined as in \cite{wu_favorable_2017}:
	\begin{align}
	\left(\mathbf{a}_r(\phi,\theta)\right)_m = e^{j \frac{\pi d_r}{\sin(\pi/M_r)} \sin{\theta}\cos{\left(\phi - \Psi_{m}\right)}},\label{eq:ar_steering_vector}
	\end{align}
	where $\Psi_m = 2\pi m/M_r$. Using this definition, followed by a cosine expansion and the simplification in \eqref{eq:HURA_steering_vector_2}, \eqref{eq:kappa_raw} requires
	\begin{align*}
	\mathbb{E}_{\phi,\theta}[e^{j\frac{\pi d_r}{\sin(\pi/M_r)}\sin\theta[\cos(\phi - \Psi_{m'})-\cos(\phi - \Psi_{m})]}]\\
	= \mathbb{E}_{\phi,\theta}[e^{j\frac{\pi d_r}{\sin(\pi/M_r)}\sqrt{a^2 + b^2}\sin\theta\sin(\phi + \tan^{-1}(b/a))}],\numberthis\label{eq:UCirA_kappa_form}
	\end{align*}
	Again, this is identical to \eqref{eq:HURA_steering_vector_2} with $a = \sin\Psi_{m'} - \sin\Psi_{m}$, $b = \cos\Psi_{m'} - \cos\Psi_{m}$, $z_1 = \pi d_r\sqrt{a^2 + b^2}/\sin(\pi/M_r)$, $z_2 = 0$, $A = \tan^{-1}(b/a)$.
	
	Both \eqref{eq:ULA_steering_vector} and \eqref{eq:UCirA_kappa_form} require the solution for \eqref{eq:HURA_steering_vector_2} derived in \cite{JSTSP}:
	\begin{align*}
		\kappa^\textrm{HURA} &=\frac{1}{M^2}\sum_{m_x,m_x'}^{M_x - 1}\sum_{m_y,m_y'}^{M_y - 1}|I(A,z_1,z_2)|^2,\numberthis\label{eq:kappa_HURA_cf}
	\end{align*}
	with
	\begin{align*} 
	I(A,z_1,z_2)&=\sum_{n=-\infty}^{\infty}\sum_{n'=-\infty}^{\infty}(-1)^{p(n)}\psi(n)\chi(2n')e^{jnA}\numberthis\label{eq:I_definition}\\
	&\times J_{\frac{|n|}{2}-n'}\left(\sqrt{\frac{z_1^2+z_2^2}{4}}\right)J_{\frac{|n|}{2}+n'}\left(\sqrt{\frac{z_1^2+z_2^2}{4}}\right)
	\end{align*}
	where $p(n) = \min(n,0)$, $J_n(\cdot)$ is the $n^\textrm{th}$ order Bessel function of the first kind, and we define $\sum_{m,m'}^{M-1} \triangleq \sum_{m=0}^{M-1}\sum_{m'=0}^{M-1}$. Here, $\psi(n) = \mathbb{E}_{\phi}[\exp(jn\phi)]$ and $\chi(n) = \mathbb{E}_{\theta}[\exp(jn\theta)]$ are the characteristic functions of the azimuth and elevation angular PDFs. In \cite{JSTSP} we show that the infinite summations in \eqref{eq:I_definition} (and later in \eqref{eq:V_definition}) can be truncated to a handful of terms while maintaining exceptional accuracy due to the rapid decay of the characteristic functions for realistic angular distributions. The NMI for a ULA and UCirA follow from \eqref{eq:kappa_HURA_cf}, as explained in Result~\ref{lemma:kappa_ULA_UCirA_cf}.
	\begin{result}\label{lemma:kappa_ULA_UCirA_cf}
		$\kappa^\textrm{ULA}$ and $\kappa^\textrm{UCirA}$ are given by the right-hand side of \eqref{eq:kappa_HURA_cf} with $M_x = 1$, $M_y = M$ and $[A,z_1,z_2]$ as in Table \ref{table:topology_parameters}. 
	\end{result}

	\subsection{UCylA, VURA}\label{ss:analysis_UCylA}
		Consider a UCylA with $M_z$ layers stacked vertically with $d_z$ wavelength spacing, each comprising a UCirA with $M_r$ antennas spaced $d_r$ apart, where $M_rM_z = M$. In this case,
		\begin{equation}\label{eq:UCylA_steering_vector}
		\mathbf{a}(\phi,\theta) = \mathbf{a}_r(\phi,\theta) \otimes \mathbf{a}_z(\phi,\theta),
		\end{equation}
		with the entries of the $M_r \times 1$ vector $\mathbf{a}_r(\phi,\theta)$ given by \eqref{eq:ar_steering_vector} and those of the $M_z \times 1$ steering vector $\mathbf{a}_z(\theta)$ defined as
		\begin{align}\label{eq:az_steering_vector}
		\left(\mathbf{a}_z(\theta)\right)_m = e^{j2\pi d_z (m-1) \cos{\theta}}.
		\end{align}
		For a UCylA with steering vectors as in \eqref{eq:UCylA_steering_vector}, \eqref{eq:kappa_raw} requires
		\begin{align*}
		\mathbb{E}_{\phi,\theta}&[e^{j\frac{\pi d_r\sin\theta}{\sin(\pi/M_r)}(\cos(\phi -\Psi_{m_r'})-\cos(\phi - \Psi_{m_r}))}\\
		&\times e^{j2\pi d_z(m_z'-m_z)\cos\theta}]\\
		&=\mathbb{E}_{\phi,\theta}[e^{jz_1\sin\theta\sin(\phi + A) + jz_2\cos\theta}]\numberthis\label{eq:UCylA_steering_vector_2}
		\end{align*}
		with $[a,b,z_1,A]$ as for a UCirA, and $z_2 = 2\pi d_z (m_z'-m_z)$.
		
		Finally, consider a VURA with $M_z$ vertically stacked rows $d_z$ apart, each having $M_y$ antennas with spacing $d_y$ parallel to the $y$-axis and $M_yM_z = M$. The steering vectors here are
		\begin{align*}
		\mathbf{a}(\phi,\theta) = \mathbf{a}_y(\phi,\theta) \otimes \mathbf{a}_z(\theta),\numberthis\label{eq:VURA_steering_vector}
		\end{align*}
		using the entries of the $M_y \times 1$ and $M_z \times 1$ vectors defined in \eqref{eq:ay_steering_vector} and \eqref{eq:az_steering_vector}. For this topology, \eqref{eq:kappa_raw} requires
		\begin{align*}
		\mathbb{E}_{\phi,\theta}[e^{j2\pi[d_y(m_y-m_y')\sin\theta\sin\phi + d_z(m_z-m_z')\cos\theta]}].\numberthis\label{eq:VURA_steering_vector_2}
		\end{align*}
		Hence, $\kappa^\textrm{VURA}$ also requires \eqref{eq:UCylA_steering_vector_2} with $z_1 = 2\pi d_y(m_y-m_y')$, $z_2 = 2\pi d_z(m_z-m_z')$, and $A = 0$.
		
		The solutions for $\kappa^\textrm{UCylA}$ and $\kappa^\textrm{VURA}$ are given in Lemma~\ref{lemma:kappa_UCylA_VURA_cf}.
		
		\begin{lemma}\label{lemma:kappa_UCylA_VURA_cf}
			For a UCylA,
			\begin{align*}
			\kappa^\textrm{UCylA} &=\frac{1}{M^2}\sum_{m_r,m_r'}^{M_r-1}\sum_{m_z,m_z'}^{M_z-1}|V(A,z_1,z_2)|^2\numberthis\label{eq:kappa_UCylA_cf}
			\end{align*}
			with
			\begin{align*}
			&V(A,z_1,z_2) = \sum_{n=-\infty}^{\infty}(-1)^{p(n)}e^{jnA}\psi(n)\sum_{n'=-\infty}^{\infty}\chi(2n')\\ &\times \sum_{\hat{n}=-\infty}^{\infty}J_{|n|/2-\hat{n}}\left(\frac{z_1}{2}\right)J_{|n|/2+\hat{n}}\left(\frac{z_1}{2}\right)G(n,n',\hat{n},z_2)\numberthis\label{eq:V_definition},
			\end{align*}
			using the previous definitions of $p(n)$, $\psi(n)$, and $\chi(n)$, and
			\begin{align*}
			&G(n,n'\hat{n},z_2) = J_{2(n'+\hat{n})}(z_2) \\
			&+ \frac{4}{\pi}\sum_{\hat{n}'=1}^{\infty}(-1)^{(\hat{n}'-n'-\hat{n})}\frac{J_{2\hat{n}'-1}(z_2)2(n'+\hat{n})}{(2\hat{n}'-1)^2-4(n'+\hat{n})^2}.\numberthis\label{eq:Gamma_definition}
			\end{align*}
			
			For a VURA, $\kappa^\textrm{VURA}$ is given by \eqref{eq:kappa_UCylA_cf} with $M_r = M_y$ and $[A,z_1,z_2]$ given in Table \ref{table:topology_parameters}.
		\end{lemma}
		
		\begin{IEEEproof}
			The proof is given in the Appendix.
		%	~\ref{app:HURA_Ks}.
		\end{IEEEproof}
		
		\begin{table}[ht]
			\caption{Parameters for topology-specific solutions for the NMI}
			\vspace{-0.3cm}
			\centering
			\begin{tabular}{|c||c|c|c|c|c|}
				\hline
				&HURA&ULA&VURA&UCylA&UCirA\\
				\hline\hline
				$a$&\multicolumn{3}{c|}{N/A}&\multicolumn{2}{c|}{$\sin(\psi_{m_r})-\sin(\psi_{m_r'})$}\\
				\hline
				$b$&\multicolumn{3}{c|}{N/A}&\multicolumn{2}{c|}{$\cos(\psi_{m_r})-\cos(\psi_{m_r'})$}\\
				\hline
				$A$&$\arctan(z_2/z_1)$&\multicolumn{2}{c|}{0}&\multicolumn{2}{c|}{$\arctan(b/a)$}\\
				\hline
				$z_1$&\multicolumn{3}{c|}{$2\pi d_y(m_y-m_y')$}&\multicolumn{2}{c|}{$\sqrt{a^2 + b^2}\pi d_r/(\sin(\pi/M_r))$}\\
				\hline
				$z_2$&$2\pi(m_x-m_x')$&$0$&\multicolumn{2}{c|}{$2\pi d_z(m_z - m_z')$}&0\\
				\hline
			\end{tabular}
			\label{table:topology_parameters}
		\end{table}
	\section{Numerical Results}\label{results}
	%
	%% trim and size all figures equally:
%	\newcommand{\myincludegraphics}{\includegraphics[trim=1cm 0.04cm 1.6cm 0.66cm, clip=true, width=1\columnwidth]}
%	\newcommand{\raisecapt}{\vspace{-0.8cm}}
%	\setlength{\textfloatsep}{5.0pt plus 1.0pt minus 4.0pt}
%	\setlength{\textfloatsep}{10.0pt plus 1.0pt minus 2.0pt}
%	\setlength{\textfloatsep}{10.0pt plus 1.0pt minus 2.0pt}
	%Hence, lower values of $\kappa^\textrm{FP}$ and $\kappa^\textrm{CH}$ are generally speaking indicative of higher performance.\newline
	This section provides a comparison of the NMI for five topologies. We consider angular distributions obtained from measurements reported in \cite{sangodoyin_cluster_2018}. Central cluster angles are Gaussian distributed in azimuth and Laplacian distributed in elevation while subray angles are Laplacian distributed in both cases. Angular distribution parameters are given in Table \ref{table:system_parameters}.
	\begin{table}[ht]
		\caption{Parameters for Numerical Results}
		\vspace{-0.3cm}
		\centering
		\label{table:system_parameters}
		\begin{tabular}{|c|c|}
			\hline
			\textbf{Parameter} & \textbf{Values}\\
			\hline\hline
%			cell radius, $r$ 										& 100 m\\
%			exclusion radius, $r_0$ 						& 10 m\\
%			median SNR value & -5 dB \\
%			pathloss exponent, $\Gamma$ 				& {3.8}\\
%			shadow fading standard deviation, $\sigma_{\mathrm{sf}}$ & {5.5 dB}\\
%			link gain reference distance, $d_0$ & 1 m\\
%			number of users, $L$ 								& 4\\
%			\hline
			cluster angle mean, $\mu_{\mathrm{c}}$ (azimuth), $\hat{\mu}_{\text{c}}$ (elevation) & $0\degree$, $90\degree$\\
			cluster angle variance, $\sigma_{\mathrm{c}}^2$ (azimuth), $\hat{\sigma}_{\mathrm{c}}^2$ (elevation) & {$(14.4\degree)^2$, $(1.9\degree)^2$}\\
			subray angle variance, $\sigma^2_{\mathrm{s}}$ (azimuth), $\hat{\sigma}^2_{\mathrm{s}}$ (elevation)  & {$(6.24\degree)^2$, $(1.37\degree)^2$}\\ [1ex]
			\hline
		\end{tabular}
	\end{table}
	We consider $L=4$ users in a cell with $r = 100\textrm{m}$, $r_0 = 10\textrm{m}$, pathloss as in Sec.~\ref{systemmodel} with $\Gamma=3.8$, $\sigma_\textrm{sf} = 5.5\textrm{dB}$, $d_0 = 1\textrm{m}$, and $A=1$. We assume cluster power is equally distributed among subrays $\beta_{c,s}^{(l)} = \beta_c^{(l)}/S$\cite{3GPP} where $\beta_c^{(l)}$ is set to exponentially decay from $\beta_1^{(l)}$ to $\beta_C^{(l)}$ with ratio $\beta_C^{(l)}/\beta_1^{(l)} = 0.1$ \cite{JSTSP}. The UL SNR $\rho$ in \eqref{eq:MMSE_SINR} is chosen such that the average received SNR at the BS, $\rho\beta^{(l)}$, has a median of $-5\textrm{dB}$.
%	\subsection{Imposing Space Constraints}\label{ss:space_constraints}

	Consider a $D/\sqrt{2} \times D/\sqrt{2}$ square in the $x$-$y$-plane, meaning the largest dimension (the diagonal) is length $D$. Under space-constraints, antennas are arranged with the maximum uniform spacing possible such that the topology's azimuth footprint fits within the space-constraint square. Table \ref{table:space_parameters} provides the resulting antenna spacing, $d$, given $D$. For simplicity, we assume $d_x = d_y = d_z = d_r = d$ within a given topology, and $M_x = M_y = M_z = M_r = \sqrt{M}$ for all topologies.
%	\begin{table}[ht]
%		\caption{Antenna spacing under space constraints}
%		\vspace{-0.3cm}
%		\centering
%		\begin{tabular}{|c||c|}
%			\hline
%			Topology&$d$\\
%			\hline\hline
%			\textbf{ULA}&$D/(M-1)$\\
%			\textbf{UCirA}&$(D/\sqrt{2})\sin(\pi/M)$\\
%			\textbf{HURA}&$D/(\sqrt{2}(\sqrt{M}-1))$\\
%			\textbf{UCylA}&$(D/\sqrt{2})\sin(\pi/\sqrt{M})$\\
%			\textbf{VURA}&$D/(\sqrt{M}-1)$\\
%			\hline
%		\end{tabular}
%		\label{table:space_parameters}
%	\end{table}

	\begin{table}[ht]
		\caption{Antenna spacing under space constraints}
		\vspace{-0.3cm}
		\centering
		\begin{tabular}{|c||c|c|c|}
			\hline
			\textbf{Topology}&ULA&UCirA&HURA\\
			\hline
			$\mathbf{d}$&$D/(M-1)$&$\frac{D}{\sqrt{2}}\sin(\frac{\pi}{M})$&$\frac{D}{\sqrt{2}}/(\sqrt{M}-1))$\\
			\hline
			\multicolumn{4}{c}{\vspace{-0.2cm}}\\
			\cline{1-3}
			\textbf{Topology}&UCylA&VURA&\multicolumn{1}{c}{}\\
			\cline{1-3}
			$\mathbf{d}$&$\frac{D}{\sqrt{2}}\sin(\frac{\pi}{\sqrt{M}})$&$D/(\sqrt{M}-1)$&\multicolumn{1}{c}{}\\
			\cline{1-3}
		\end{tabular}
		\label{table:space_parameters}
	\end{table}

	\begin{figure}[ht]
		\centering
%		\myincludegraphics
		\includegraphics[trim=1cm 0.04cm 1.6cm 0.1cm, clip=true, width=1\columnwidth]  
		{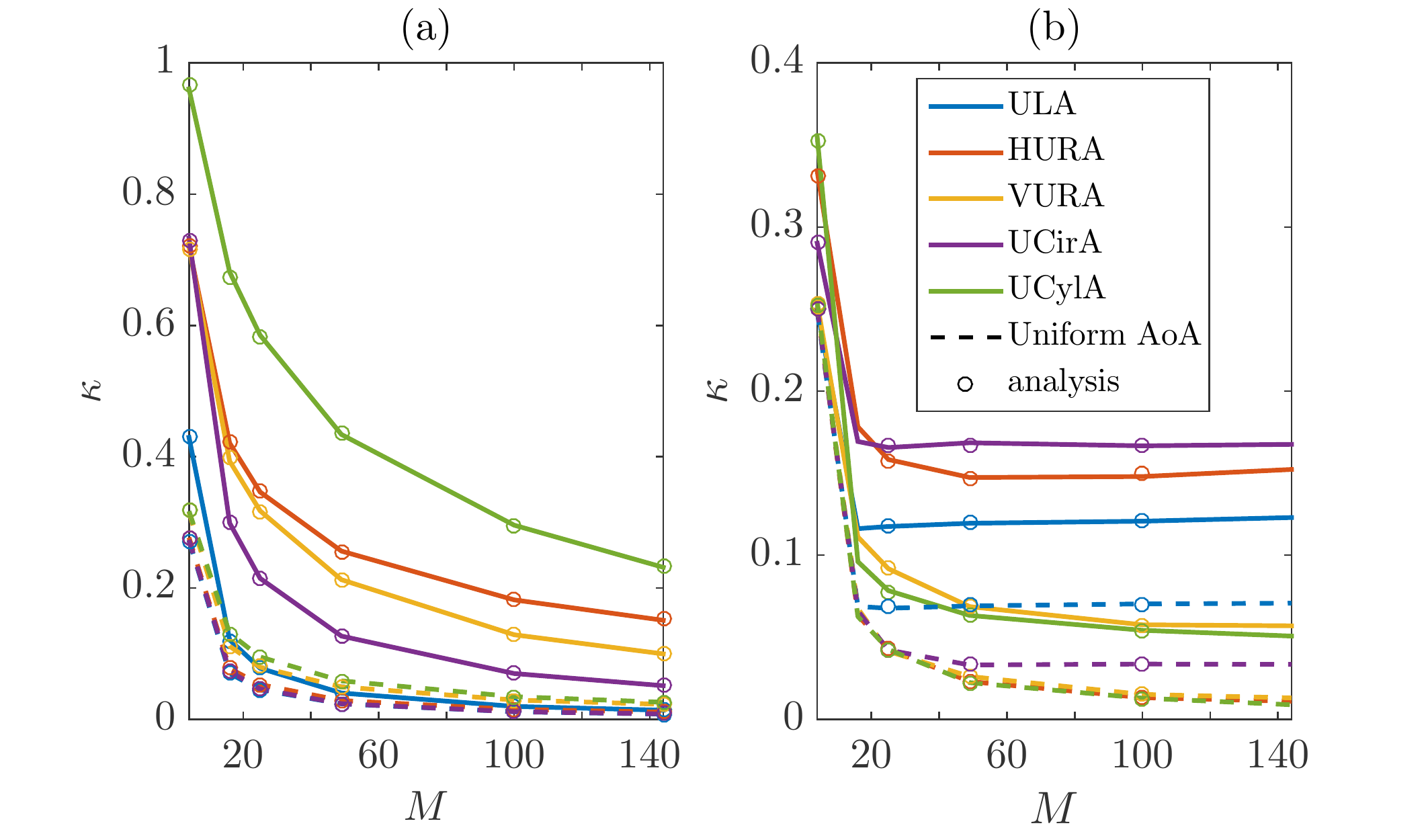}
%		\raisecapt
		\caption{NMI vs $M$ for: (a) unconstrained arrays with $d=0.5$; (b) constrained arrays with $D=7.77$.}
		\label{fig:kappa_vs_M}
	\end{figure}
	Fig.~\ref{fig:kappa_vs_M}(a) examines the system without space constraints by plotting the NMI for each topology vs $M$ for $d=0.5$. In Fig.~\ref{fig:kappa_vs_M}(b) we observe the NMI under space constraints with $D$ equal to that of a 144-antenna HURA with $d = 0.5$. In addition to the angular distributions from Table~\ref{table:system_parameters}, we include results for the case of AAoAs and EAoAs independently uniformly distributed on $[-\pi,\pi]$, as assumed in \cite{wu_favorable_2017,masouros_space-constrained_2015,DanielAltamirano2016}.

		First, we note that Fig.~\ref{fig:kappa_vs_M} makes it very clear that the choice of angular distributions is very important in topology comparisons. In both Fig.~\ref{fig:kappa_vs_M}(a) and Fig.~\ref{fig:kappa_vs_M}(b) the uniform AAoAs and EAoAs give drastically lower values of the NMI and reveal little to no topology-specific trends. This is the most spatially diverse angular distribution, hence any topology-specific characteristics which promote spatial diversity under less diverse conditions will be obscured in the uniform case. In Fig.~\ref{fig:kappa_vs_M}(b), only the ULA and UCirA have noticeably larger NMI values, as they have the smallest inter-element spacing under space constraints (see Table \ref{table:space_parameters}). The ULA additionally suffers from increased endfire radiation in a uniform AoA distribution (see \cite{chelsea_icc19}, \cite{JSTSP}).

	Secondly, the considerable difference in NMI trends between Fig.~\ref{fig:kappa_vs_M}(a) and Fig.~\ref{fig:kappa_vs_M}(b) speaks to the importance of observing topology behaviour under space constraints. In Fig.~\ref{fig:kappa_vs_M}(a), the topology-specific benefits are obscured by the significant advantage afforded to those with larger azimuthal apertures; only once this advantage is removed can the effects of antenna arrangement be observed.

	For measurement-based angular distributions in Fig.~\ref{fig:kappa_vs_M}(a), the level of interference and correlation decrease in the following order: UCylA, HURA, VURA, UCirA, ULA. The azimuthal apertures increase in this order, showing how greatly this affects performance without space constraints\cite{aslam_performance_2019}. As expected, once the same space constraint is applied  in Fig.~\ref{fig:kappa_vs_M}(b), the differences between the topologies decrease. However, the ordering of topologies is now completely changed,  decreasing in the following order:  UCirA, HURA, ULA, VURA, UCylA. Hence, the topologies with elevation variation now dominate.
	%This pattern is broken only by the lower values of the NMI for a VURA as compared to an HURA. Although the maximum azimuthal aperture is a factor of $\sqrt{2}$ greater for the latter, in this case the increased elevation resolution afforded by the antenna placement of the VURA provides greater diversity gain than the larger azimuthal aperture of the HURA.
	
\begin{figure}[ht]
		\centering
		%		\myincludegraphics
		\includegraphics[trim=1cm 1.3cm 1.6cm 0.1cm, clip=true, width=1\columnwidth] 
		{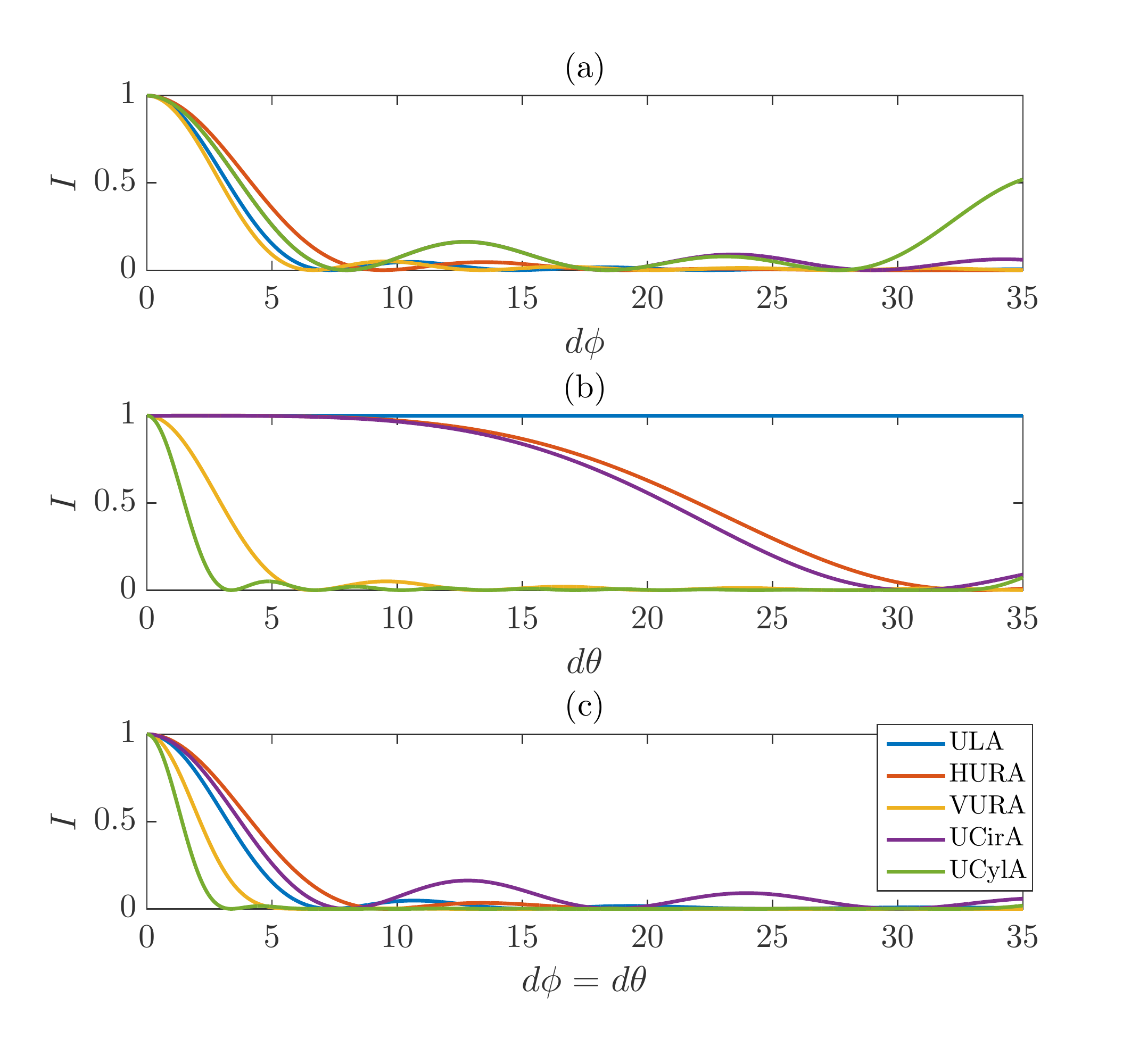}
		%		\raisecapt
		\caption{Normalized interference power of two rays separated by $d\phi^o$ and/or $d\theta^o$, $M = 100$, $D = 7.77$. For all, $\{\phi_1, \theta_1\} = \{0,\pi/2\}$. The second ray is separated in azimuth and elevation as (top to bottom): $\{\phi_2, \theta_2\} = \{\phi_1 + d\phi, \theta_1\}$; $\{\phi_2, \theta_2\} = \{\phi_1, \theta_1 + d\theta\}$; $\{\phi_2, \theta_2\} = \{\phi_1 + d\phi, \theta_1 + d\theta\}$.}
		\label{fig:inter_vs_delta_az_el}
	\end{figure}

	To assist in explaining this marked change in the ordering of space-constrained topologies, in Fig.~\ref{fig:inter_vs_delta_az_el} we observe the normalised interference power of two rays, $I = \mathbb{E}_{\Theta}[|\mathbf{h}_1^\textrm{H}\mathbf{h}_2|^2/(\beta_1\beta_2)] = |\mathbf{a}(\phi_1,\theta_1)^\textrm{H}\mathbf{a}(\phi_2,\theta_2)|^2$. One ray approaches at broadside, and the other at some offset angle $d\phi$ in azimuth and/or $d\theta$ in elevation. This illustrates the \textit{angular resolution}, where lower interference implies greater resolution.
	
	 Fig.~\ref{fig:inter_vs_delta_az_el}(a) illustrates that all topologies exhibit reasonably similar azimuth resolution. %, with the exception of the UCirA and UCylA, for which the azimuth resolution fluctuates more significantly for larger ray separations. 
	 The elevation resolution in Fig.~\ref{fig:inter_vs_delta_az_el}(b) paints an entirely different picture. The horizontal configurations of the ULA, HURA and UCirA require significantly larger angular separation in elevation than in azimuth to achieve the same interference reduction. Most measurements, including \cite{sangodoyin_cluster_2018}, suggest a reasonably small elevation spread so that such separations are less likely. In contrast, the VURA and UCylA achieve good elevation resolution with only a few degrees of separation, explaining the dominance of vertical topologies under space constraints. Fig.~\ref{fig:inter_vs_delta_az_el}(c), shows the resolution when the rays are separated by a common angle  $d\phi=d\theta$ in azimuth and elevation, hence capturing the trends in both Fig.~\ref{fig:inter_vs_delta_az_el}(a) and Fig.~\ref{fig:inter_vs_delta_az_el}(b). Note that the initial rate of decay out to $8^o$ separation follows almost the same ordering as the NMI shown in Fig.~\ref{fig:kappa_vs_M}(b). The only difference is that the HURA/UCirA ordering is reversed due to the ripples present at wider separations. In summary, the broad behaviour is that a fixed azimuth footprint reduces the azimuthal differences between topologies  so that elevation resolution is the dominant factor, despite the lesser angular variation in elevation.

	\begin{figure}[ht]
		\centering
%		\myincludegraphics
		\includegraphics[trim=1cm 1.3cm 1.6cm 0.5cm, clip=true, width=1\columnwidth] 
		{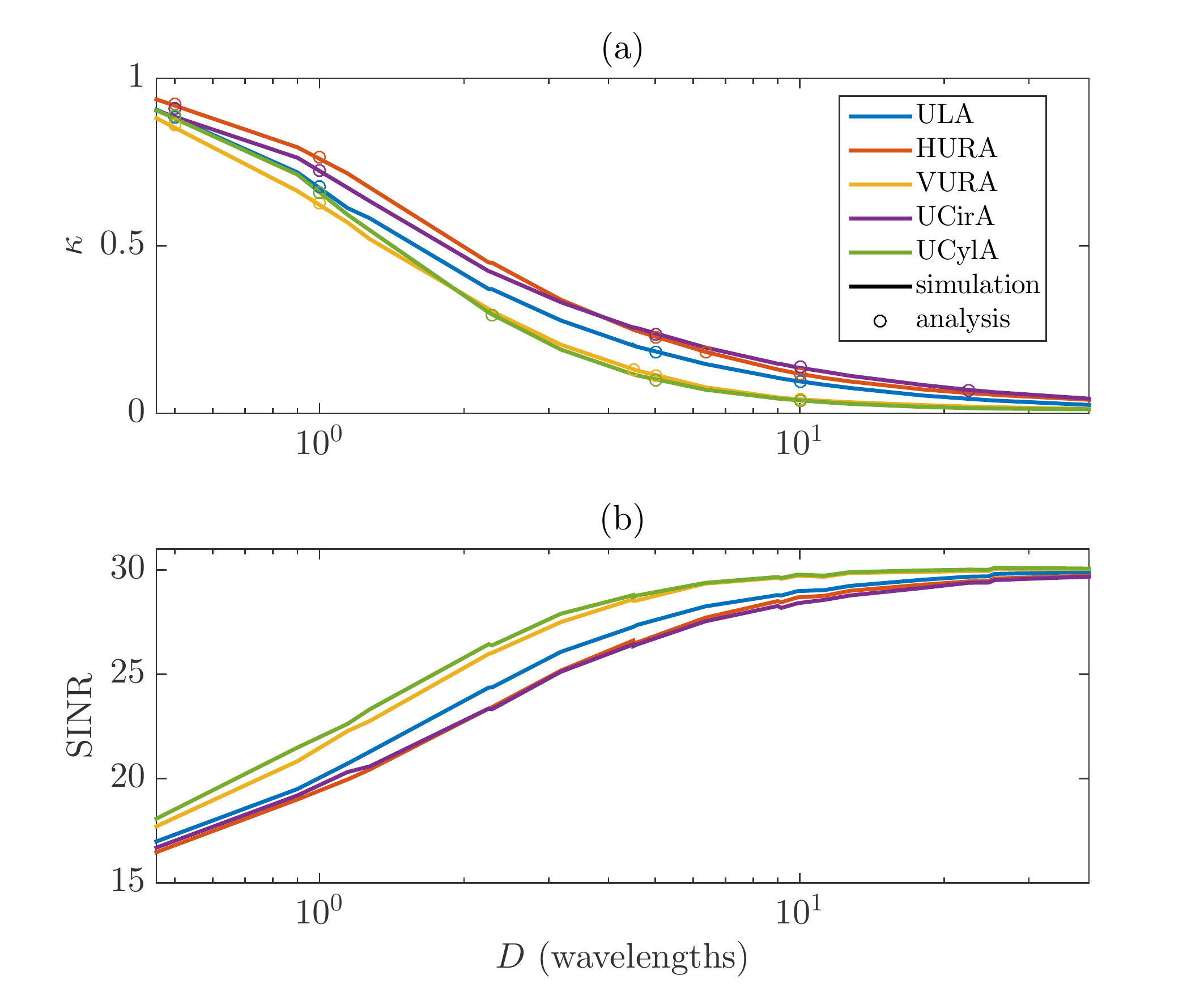}
%		\raisecapt
		\caption{(a) NMI and (b) MMSE SINR vs $D$, $M=100$.}
		\label{fig:kappa_vs_D_constant_M}
	\end{figure}

	In Fig.~\ref{fig:kappa_vs_D_constant_M}(a) we compare the NMI for a fixed number of antennas across a wide range of $D$, while in Fig.~\ref{fig:kappa_vs_D_constant_M}(b) we examine the per-user cell-wide average SINR with MMSE processing using \eqref{eq:MMSE_SINR}. Fig.~\ref{fig:kappa_vs_D_constant_M}(a) shows that the pattern observed in Fig.~\ref{fig:kappa_vs_M}(b) is roughly consistent regardless of total size until D becomes large, when the NMI values for all topologies converge near zero. This implies that, given a large enough BS aperture, the configuration of the antennas becomes unimportant. This is encouraging for concepts such as distributed MIMO where the BS aperture is very large, but antenna placement might be limited. The pattern displayed in the ergodic per-user SINR with MMSE processing agrees with the NMI ordering (as the NMI drops the SINR increases) validating the NMI as a measure of relative performance.
%	\begin{figure}[ht]
%	\centering
%	%		\myincludegraphics
%	\includegraphics[trim=1cm 0.04cm 1.6cm 0.78cm, clip=true, width=1\columnwidth] 
%	{interference_power_vs_ray_separation_azimuth.pdf}
%	%		\raisecapt
%	\caption{SINR vs $D$ for all antenna topologies: $M=100$. Vertical colored lines indicate the value of $D$ at which the corresponding topology has half-wavelength antenna spacing}
%	\label{fig:inter_vs_delta_az}
%	\end{figure}

%	\begin{figure}[ht]
%	\centering
%	%		\myincludegraphics
%	\includegraphics[trim=1cm 0.04cm 1.6cm 0.78cm, clip=true, width=1\columnwidth] 
%	{interference_power_vs_ray_separation_elevation.pdf}
%	%		\raisecapt
%	\caption{SINR vs $D$ for all antenna topologies: $M=100$. Vertical colored lines indicate the value of $D$ at which the corresponding topology has half-wavelength antenna spacing}
%	\label{fig:inter_vs_delta_el}
%	\end{figure}

%	\begin{figure}[ht]
%	\centering
%	%		\myincludegraphics
%	\includegraphics[trim=1cm 0.04cm 1.6cm 0.78cm, clip=true, width=1\columnwidth] 
%	{SINR_vs_bigD_lowspread_M100_sim_only.pdf}
%	%		\raisecapt
%	\caption{SE vs $D$ for all antenna topologies with $\rho = 10dB$: $M=100$. Vertical colored lines indicate the value of $D$ at which the corresponding topology has half-wavelength antenna spacing.}
%	
%	\label{fig:SINR_vs_M_constant_D}
%	\end{figure}

	%
	\section{Conclusions}\label{conclusion}
	We provide closed-form equations for the NMI for five antenna topologies, which we use to study their behaviour with and without space constraints. For fixed antenna spacing, topologies with \textit{wider azimuth footprints} are advantageous. Under space constraints, the NMI is determined by the topology's \textit{angular resolution}, particularly in elevation. SINR trends with MMSE processing confirm these conclusions.
	%
	%
	%
	%%%%%%%%%%%% APPENDIX %%%%%%%%%%%%%%%
	%
	\appendix[Proof of Lemma \ref{lemma:kappa_UCylA_VURA_cf}]\label{app:kappa_VURA}
	From \eqref{eq:UCylA_steering_vector_2}, we require an expectation of the form
	\begin{align*}
	V(A,z_1,z_2) = \mathbb{E}_{\theta}[\hspace{0.2cm}&\mathbb{E}_{\phi}[\exp(j[z_1\sin\theta\sin(\phi + A)])]\\
	&\times\exp(jz_2\cos\theta)].\numberthis\label{eq:app_UCylA_form}
	\end{align*}
	Using the analysis from \cite{JSTSP}, \eqref{eq:app_UCylA_form} becomes
	\begin{align*}
	&\sum_{n=-\infty}^{\infty}\psi(n)e^{jnA}\mathbb{E}_{\theta}[J_n(z_1\sin\theta)\exp(jz_2\cos\theta)]\\
	&= \sum_{n=-\infty}^{\infty}\sum_{n' = -\infty}^{\infty}\psi(n)e^{jnA}(-1)^{n'}\chi(2n')\\
	&\times\frac{1}{\pi}\int_{0}^{\pi}J_n(z_1\sin\theta)\exp(j[z_2\cos\theta-2n'\theta])d\theta.\numberthis\label{eq:app_UCylA_form_2}
	\end{align*}
	We substitute the Bessel function in \eqref{eq:app_UCylA_form_2} with its Fourier series and use \cite[Equations 6.681.8 and 6.681.9]{gradshteyn2007} to give
	\begin{align*}
	&\int_{0}^{\pi}J_n(z_1\sin\theta)\exp(j[z_2\cos\theta-2n'\theta])d\theta\\
	&= 
	\sum_{\hat{n}=-\infty}^{\infty}(-1)^{\hat{n}+p(n)}J_{|n|/2-\hat{n}}(z_1/2)J_{|n|/2+\hat{n}}(z_1/2)\\
	&\times\int_{0}^{\pi}\exp(j[z_2\cos\theta-2(n'+\hat{n})\theta])d\theta.\numberthis\label{eq:app_besselexpansion}
	\end{align*}
	We now require $\int_{0}^{\pi}\exp(j[z_2\cos\theta-z_3\theta])d\theta$ with $z_3 = 2(n'+\hat{n})$. Expanding the exponential and using the symmetry of $\sin(z_2\sin - z_3\theta)$ around $\pi$, this becomes
	\begin{align*}
	(-1)^{z_3/2}\bigg[&\int_{\pi/2}^{3\pi/2}\cos(z_2\sin\theta)\cos(z_3\theta)d\theta +\\
	&\int_{\pi/2}^{3\pi/2}\sin(z_2\sin\theta)\sin(z_3\theta)d\theta\bigg].\numberthis\label{eq:app_expcosintegral_2}
	\end{align*}
	Using \cite[Eq. 8.411.2]{gradshteyn2007}, we have
	\begin{align*}
	\int_{\pi/2}^{3\pi/2}\cos(z_2\sin\theta)\cos(z_3\theta)d\theta = \pi J_{z_3}(z_2).\numberthis\label{eq:app_cossinintegral}
	\end{align*}
	For the remaining integral, \cite[Eq. 8.514.6]{gradshteyn2007} provides
	\begin{align*}
	&\sin(z_2\sin\theta)\sin(z_3\theta)\hspace{-0.1cm}=\hspace{-0.1cm}2\hspace{-0.1cm}\sum_{\hat{n}'=1}^{\infty}\hspace{-0.15cm}J_{2\hat{n}'-1}(z_2)\sin((2\hat{n}'\hspace{-0.1cm}-\hspace{-0.1cm}1)\theta)\sin(z_3\theta).
	\end{align*}
	A cumbersome but straightforward progression therefore gives
	\begin{align*}
	&\int_{\pi/2}^{3\pi/2}\sin(z_2\sin\theta)\sin(z_3\theta)d\theta =\\ &2\sum_{\hat{n}'=1}^{\infty}J_{2\hat{n}'-1}(z_2)(-1)^{z_3/2+\hat{n}'}\frac{2z_3}{(2\hat{n}'-1)^2-z_3^2}.\numberthis\label{eq:app_sinsinintegral}
	\end{align*}
	Substituting \eqref{eq:app_cossinintegral} and \eqref{eq:app_sinsinintegral} into \eqref{eq:app_expcosintegral_2}, we have
	\begin{align*}
	&\int_{0}^{\pi}\exp(j[z_2\cos\theta-z_3\theta])d\theta = (-1)^{z_3/2}\pi J_{z_3}(z_2) +\\ &4\sum_{\hat{n}'=1}^{\infty}(-1)^{\hat{n}'}J_{2\hat{n}'-1}(z_2)\frac{z_3}{(2\hat{n}'-1)^2-z_3^2}.\numberthis\label{eq:app_expcosintegral_cf}
	\end{align*}
	Finally, substituting \eqref{eq:app_expcosintegral_cf} into \eqref{eq:app_UCylA_form_2} gives the solution in \eqref{eq:V_definition}.

	\bibliographystyle{IEEEtran}
	\bibliography{bibliography}
%	\bibliography{bibliography}
	%
\end{document}